\documentclass[sigconf, nonacm]{acmart}

\usepackage[utf8]{inputenc}  
\usepackage{float}           
\usepackage{multirow}

\title{MoE-MLoRA for Multi-Domain CTR Prediction: Efficient Adaptation with Expert Specialization } 

\author{Ken Yaggel$^*$}
\author{Eyal German$^*$}
\author{Aviel Ben Siman Tov$^*$}
\affiliation{%
  \institution{Department of Software and Information Systems Engineering, Ben-Gurion University of the Negev}
  \city{Be'er Sheva}
  \country{Israel}
}
\email{{kenyag, germane, avielben}@post.bgu.ac.il}

\thanks{$^*$Equal contribution.}



\begin{document}

\begin{abstract}
Personalized recommendation systems must adapt to user interactions across different domains. 
Traditional approaches like MLoRA apply a single adaptation per domain but lack flexibility in handling diverse user behaviors. 
To address this, we propose MoE-MLoRA, a mixture-of-experts framework where each expert is first trained independently to specialize in its domain before a gating network is trained to weight their contributions dynamically. 
We evaluate MoE-MLoRA across eight CTR models on Movielens and Taobao, showing that it improves performance in large-scale, dynamic datasets (+1.45 Weighed-AUC in Taobao-20) but offers limited benefits in structured datasets with low domain diversity and sparsity.
Further analysis of the number of experts per domain reveals that larger ensembles do not always improve performance, indicating the need for model-aware tuning.
Our findings highlight the potential of expert-based architectures for multi-domain recommendation systems, demonstrating that task-aware specialization and adaptive gating can enhance predictive accuracy in complex environments.
The implementation and code are available in our \href{https://github.com/Kenyaggel/MLoRA}{GitHub repository}.

\end{abstract}

\maketitle

\section{Introduction}
\label{sec:introduction}

Click-Through Rate (CTR) prediction is a fundamental task in recommendation systems, aiming to estimate the probability that a user will click on a given item~\cite{10.5555/3157382.3157473, 7837964, 10.1145/2959100.2959134}. Accurate CTR prediction is crucial for optimizing user engagement and maximizing revenue in various online platforms, such as e-commerce, streaming services, and digital advertising.

Traditional CTR prediction models are often designed for single-domain applications, where a model is trained on a specific dataset with consistent feature distributions. However, in real-world scenarios, recommendation systems frequently operate across multiple domains, each with distinct user behaviors, item distributions, and contextual dependencies. This introduces challenges, such as domain shift, data sparsity, and feature inconsistency, which can degrade model performance.

To address these challenges, researchers have explored multi-domain CTR prediction strategies that balance cross-domain knowledge sharing with domain-specific adaptation. LoRA~\cite{hu2022lora} is a parameter-efficient fine-tuning technique that enables adapting large-scale models with minimal computational overhead. MLoRA~\cite{mlora_recsys2024} extends this approach for multi-domain CTR prediction by dynamically sharing LoRA parameters across domains while preserving domain-specific representations. Unlike traditional LoRA, which typically trains separate fine-tuned models per domain, MLoRA leverages modularized adapters to enhance scalability and cross-domain generalization.
However, this approach restricts each domain to a single fine-tuned adapter, potentially limiting the effective transfer of shared knowledge across domains.


In this work, we propose a novel approach that integrates Mixture-of-Experts (MoE) with LoRA to enhance multi-domain CTR prediction. Our method, \textit{MoE-MLoRA}, treats each LoRA adapter as an expert and introduces a gating function to dynamically aggregate knowledge from multiple experts. 
This design enables more effective parameter sharing, enhances domain specialization, and improves cross-domain generalization.

We propose a three-stage training process: (1) training the backbone model without LoRA adapters, (2) fine-tuning each domain-specific expert separately while freezing the rest of the model, and (3) training a gating function across all domains to optimize expert selection dynamically. We demonstrate that our approach achieves state-of-the-art performance in multi-domain CTR prediction while maintaining a low computational cost.


\section{Background and Related Work}
\label{sec:related_work}
This section explores key developments in CTR prediction, multi-domain modeling, and parameter-efficient adaptation, including Mixture of Experts (MoE) and LoRA-based methods.

\subsection{CTR Prediction}
Click-Through Rate (CTR) prediction is a fundamental task in recommendation systems, aiming to estimate the probability of a user clicking on a given item. Traditional CTR models are often trained for a single domain, limiting their ability to generalize across multiple domains with varying user behaviors, content distributions, and contextual dependencies.

One approach to addressing this challenge is through deep learning models specifically adapted for CTR prediction. These models typically consist of an embedding layer followed by multi-layer perceptron (MLP) layers. The embedding layer transforms sparse categorical features into dense vectors, which are then processed by the MLP to learn complex feature interactions.

One well-known method is Wide\&Deep Learning~\cite{wide_and_deep}, which combines a linear model for memorization with an MLP for generalization. Similarly, DeepFM~\cite{guo2017deepfmfactorizationmachinebasedneural} integrates a Factorization Machine into the architecture to enhance feature interactions.
Other approaches, such as AutoInt~\cite{AutoInt} and CAN~\cite{bian2021canfeaturecoactionclickthrough}, leverage self-attention mechanisms to capture intricate dependencies between features, enabling more comprehensive feature interactions.

\subsubsection*{Multi-Domain CTR Prediction}
Recent research has focused on multi-domain CTR prediction, where a single model is designed to handle data from multiple domains efficiently. This setting presents challenges such as domain shift, data sparsity, and scalability. 
Domain shift occurs when user behavior, item distributions, and interaction patterns vary across domains, making it difficult for a single model to generalize effectively. Data sparsity poses another challenge, as some domains may have limited interaction data, leading to biased or suboptimal model training. Additionally, scalability becomes a concern as maintaining separate models for each domain is computationally expensive, requiring efficient parameter sharing while preserving domain-specific adaptability. 

One approach to addressing these challenges is training a shared model on data from multiple domains~\cite{dredze2010multi, li2020improving}. This strategy helps mitigate data sparsity to some extent but struggles to capture domain-specific nuances and variations in data distribution.

An improved solution involves leveraging Domain Generalization (DG) methods~\cite{9847099}, which extract common knowledge from multiple domains and learn generalizable features that transfer effectively to unseen domains~\cite{Li_2017_ICCV}. By focusing on shared knowledge extraction, DG techniques help alleviate data sparsity while enhancing model robustness.

To address these challenges more effectively, researchers have proposed hybrid architectures that incorporate both shared and domain-specific components. These architectures learn an overall data distribution through shared layers while introducing domain-specific layers to capture unique characteristics of each domain.
One such method is the Progressive Layered Extraction (PLE) framework~\cite{10.1145/3383313.3412236}, which enables joint learning from multiple domains by explicitly separating shared and task-specific components. However, a key limitation of PLE is that its task-specific components closely resemble the shared parts, leading to a substantial increase in model parameters and computational costs.


\subsection{MLoRA}
In the MLoRA paper~\cite{mlora_recsys2024}, they proposed a new technique for CTR prediction in a multi-domain setting. 
Their method is based on LoRA~\cite{hu2022lora}, a fine-tuning approach that reduces memory requirements during the fine-tuning process by introducing low-rank adaptation layers. 
Unlike traditional fine-tuning methods that require updating a large number of parameters, LoRA injects small, trainable rank decomposition matrices into each layer of the model, thereby efficiently adapting pre-trained models to new tasks.

MLoRA extends this concept by incorporating domain-specific adapters that enable efficient adaptation across multiple domains while preserving shared knowledge. Their approach ensures that the model retains generalization capabilities across different domains while reducing computational overhead. They demonstrate that MLoRA outperforms existing methods in multi-domain CTR prediction while maintaining low computational costs compared to full fine-tuning approaches. However, while their method is efficient and achieves strong performance, we argue that it does not fully leverage the knowledge captured by the adapters and makes no explicit attempt to transfer information between domains beyond what is learned by the main network.

\subsection{LoRA-MoE methods}
Mixture of Experts (MoE) improves efficiency and specialization by distributing computation across multiple experts. 
Instead of using a shared model for all tasks, MoE assigns different experts to different tasks, enabling better task-specific learning. 
Several works have explored the combination of MoE and LoRA for parameter-efficient fine-tuning. One such approach integrates MoE with LoRA for multi-task medical applications, including named entity recognition and diagnosis report generation~\cite{liu2023moelora}. This method employs a learned gating function to dynamically assign weights to experts based on the input task, enabling adaptive knowledge transfer while maintaining efficiency. By jointly training experts and the gating function, this framework improves task specialization and outperforms both standard LoRA and full fine-tuning strategies.

Another study applies MoE to LoRA by treating LoRA modules as distinct experts and selectively activating only a subset per input~\cite{luo2024moelora}. This design reduces computational costs while enhancing expert specialization using contrastive learning, which encourages experts to learn distinct features rather than redundant information. The approach has demonstrated effectiveness in reasoning tasks, showing notable improvements in both mathematical and common-sense reasoning over standard LoRA.

Unlike previous approaches that train all experts simultaneously, our method adopts a sequential training paradigm to improve specialization and adaptability. Each expert is first trained independently on its respective domain and then frozen, ensuring that domain-specific knowledge is well-preserved. Once the experts are established, we use a learned gating function that dynamically assigns weights across domains, leveraging cross-domain relationships without resorting to sparse expert activation. 

\section{Method}
\label{sec:method}
\begin{figure*}[t]
    \centering
    \includegraphics[width=1\textwidth]{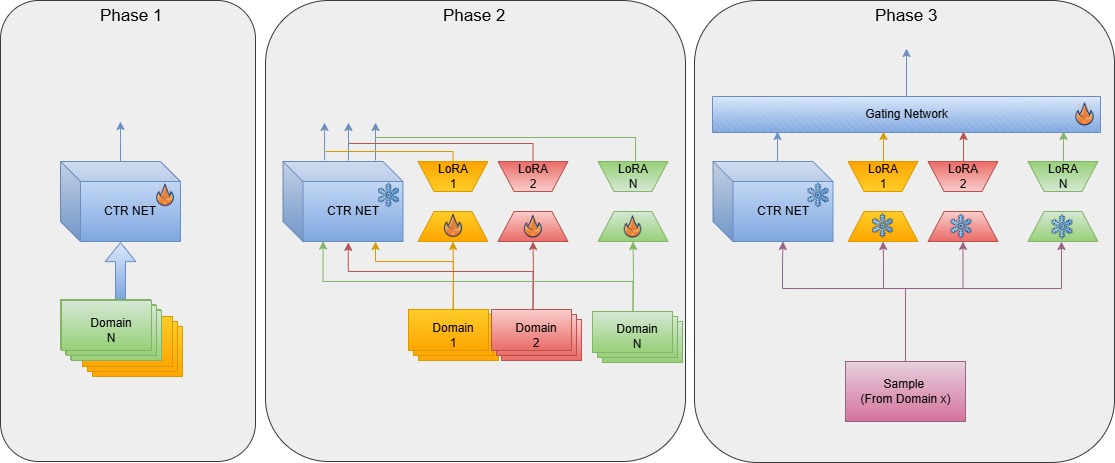} 
    \caption{Overview of the structured three-phase training process. The workflow illustrates the integration of MoE with LoRA, depicting the sequential phases: (1) Backbone Pre-training, (2) Domain-Specific Fine-Tuning, and (3) Gating Optimization. Each phase is designed to balance domain specialization and cross-domain adaptability.}
    \label{fig:method}
\end{figure*}

Our proposed method replaces the original \textit{MLoRA} layers in each model with the \textit{MloraMoE} layer, introducing a more structured mixture-of-experts (MoE) approach to multi-domain CTR prediction. 
Unlike traditional MoE frameworks that train all experts simultaneously, our method enforces domain specialization through a structured three-phase training process.
The overall training pipeline, illustrated in Figure~\ref{fig:method}, consists of the following phases:
\begin{itemize}
\item \textbf{Phase 1 - Backbone Pre-training:} First, we train only the base model to establish a shared representation across all domains. This step ensures that the model learns generalizable features before domain-specific specialization occurs.
\item \textbf{Phase 2 - Domain-Specific Fine-Tuning:} Each expert is trained separately on its respective domain while keeping the rest of the model frozen. 
Additionally, all other expert modules are excluded inference, ensuring that only the backbone model and the active expert contribute to the prediction. 
This step enforces strong domain specialization without interference from other domains.  
\item \textbf{Phase 3 - Gating Optimization:} After all expert modules have been trained, we introduce a gating function that dynamically assigns mixture weights based on input characteristics.
In this phase, all trained experts are utilized together in a frozen state, and the gating function learns to leverage cross-domain relationships while maintaining domain specialization.
\end{itemize}

This structured three-phase training approach ensures that each expert achieves strong domain expertise, while the gating function effectively blends their contributions for improved multi-domain adaptation. 
During inference, a given sample (user, item, domain) passes through both the backbone model and all expert modules. The gating function then aggregates the outputs from the backbone and experts, using the domain as an additional input to compute a weighted sum. This final aggregated output serves as the model’s prediction, effectively combining shared and domain-specific knowledge for optimal performance.


\section{Evaluation}
\label{sec:evaluation}

To assess the effectiveness of our proposed framework, we conduct a series of experiments focusing on its impact on multi-domain CTR prediction. 
We compare our approach against an existing MLoRA solution 
and investigate whether an ensemble of experts per domain provides additional benefits, compared to an individual expert.

Our experiments utilize benchmark datasets for multi-domain recommendation, derived from two publicly available real-world datasets: Taobao~\cite{du2019sequential} and Movielens~\cite{he2016ups}. To construct multi-domain datasets, we follow the domain-splitting methodology used in the MLoRA~\cite{mlora_recsys2024} paper. Specifically:
\begin{itemize}
    \item \textbf{Taobao}: This dataset contains click logs categorized by various themes. We randomly select 10 and 20 themes, treating each theme as a distinct domain.
    \item \textbf{Movielens}: This dataset contains 1 million movie ratings, which we split in three ways:
    \begin{enumerate}
        \item \textbf{By gender}: Split into two domains based on user gender.
        \item \textbf{By age}: Split into seven domains based on user age groups.
        \item \textbf{By occupation}: Split into 21 domains according to user occupation.
    \end{enumerate}
\end{itemize}


We evaluate our proposed method across eight state-of-the-art recommendation models: Wide \& Deep Learning (WDL)~\cite{wide_and_deep}, Neural Factorization Machines (NFM)~\cite{he2017neural}, AutoInt~\cite{AutoInt}, Product-based Neural Networks (PNN)~\cite{7837964}, Deep \& Cross Network (DCN)~\cite{wang2017deep}, FiBiNET~\cite{huang2019fibinet}, DeepFM~\cite{guo2017deepfmfactorizationmachinebasedneural}, and xDeepFM~\cite{lian2018xdeepfm}. The performance of each model is assessed to determine the benefits of our approach in multi-domain settings.
To compare predictive performance, we report the \textit{weighted AUC} (WAUC), computed as:

\begin{equation}\label{eqn-10} 
WAUC = \sum_{i=1}^{n} \omega_i \cdot AUC_i,
\end{equation}

where $AUC_i$ is the AUC score for the $i$-th domain, and $\omega_i$ is its relative data proportion.


\subsection{Comparison with MLoRA}
In this experiment we aim to evaluate the overall performance of our framework compared to MLoRA method. 
While MLoRA applies a single LoRA adaptation per task, our method introduces MoE with a structured training approach, where each expert is trained separately before freezing and integrating them with a gating function which weights the dominance of each expert for specific task. 
This experiment tests whether specialized domain experts with adaptive weighting provide better predictive accuracy than a single shared LoRA adapter per domain.

For this comparison, both models are trained under identical conditions, ensuring the same backbone architecture, dataset, and training procedure. 
We report the WAUC scores for each approach and quantify the improvements in predictive performance across multiple domains.



\subsection{Effect of Ensemble of Experts per Domain}
In this experiment, we investigate whether each domain benefits from an ensemble of experts rather than relying on a single specialized expert. While our primary approach assigns one expert per domain, we extend this setup by training multiple experts per domain and analyzing whether aggregating their predictions enhances accuracy.

For this analysis, we use the Movielens dataset, where the data is split by gender into two domains. We evaluate scenarios with different ensemble sizes: two experts in total (one per domain) and configurations with 4, 6, and 8 experts, corresponding to 2, 3, and 4 experts per domain, respectively. All other hyperparameters remain fixed to isolate the effect of ensemble size. The gating function is adjusted accordingly to distribute weights among the available experts.


\section{Results}
\label{sec:results}
In this section, we present our results across various models and datasets, comparing our approach to the MLoRA method.

\subsection{Comparison with MLoRA}
Table~\ref{table:main_exp_result} presents the WAUC (\%) results comparing MoE-MLoRA to the baseline MLoRA across multiple CTR models on Movielens and Taobao datasets, comparing our proposed MoE-MLoRA against the baseline MLoRA. Each model is evaluated across multiple domain splits, and improvements ($\Delta$) are reported as MoE-MLoRA minus MLoRA.


On the Movielens dataset, MoE-MLoRA demonstrates its strongest performance in the occupation-based split, achieving consistent improvements over MLoRA in most models, with an average $\Delta = +0.11$. This setting involves a larger number of domains (21 occupations), which likely introduces greater behavioral diversity and enables more effective expert specialization. In contrast, the gender and age splits, which define only two and seven domains respectively, show minimal or slightly negative gains ($\Delta = -0.14$ and $-0.10$), suggesting that when domain differences are limited, the advantages of expert-based modeling diminish. These findings highlight that MoE-MLoRA is particularly effective when applied to granular domain partitions that expose meaningful variation across user groups.

In contrast, MoE-MLoRA consistently outperforms MLoRA on both Taobao splits, with average gains of +1.43 on the 10-theme split and +1.45 on the 20-theme split.
One plausible explanation lies in the sparsity and diversity of the Taobao dataset. For example, the overall sparsity in Taobao-10 and Taobao-20 is 0.9989 and 0.9995, respectively, substantially higher than the 0.9553 sparsity observed in Movielens. 
Sparser domains present a more challenging modeling problem due to limited user-item interactions, which can make it harder for traditional models to generalize effectively. Our results suggest that MoE-MLoRA demonstrates strong advantages in such sparse and fragmented settings, where the ability to route inputs to specialized experts helps capture fine-grained, domain-specific patterns. In contrast, the more homogeneous and denser structure of Movielens—especially in the gender and age splits—may reduce the need for such specialization, limiting the added value of expert-based routing.

To ensure fair contribution across datasets, we compute an overall average improvement by averaging the per-dataset averages, resulting in a final Avg($\Delta$) = +0.70. This confirms that MoE-MLoRA provides clear benefits in complex, multi-domain scenarios such as Taobao, while remaining competitive on simpler or more homogeneous splits like Movielens.

\begin{table*}
    \centering
    \small 
    \setlength{\tabcolsep}{4pt} 
    \caption{WAUC ($\%$) results of CTR prediction on Movielens and Taobao datasets across different models. MLoRA serves as the baseline method, while MoE-MLoRA extends it by assigning one expert per layer for each domain. The $\Delta$ rows report the per-model performance difference of MoE-MLoRA compared to MLoRA (i.e., MoE-MLoRA $-$ MLoRA). \textit{Avg($\Delta$)} represents the average of these differences across all domains, computed separately for each dataset and then averaged to ensure equal contribution regardless of the number of domain splits.}
    \label{table:main_exp_result}
    \resizebox{\textwidth}{!}{
    \begin{tabular}{@{}c|c|c|c|c|c|c|c|c|c|c@{}}
    \toprule
    \textbf{Dataset} & \textbf{Approach} & WDL & NFM & AutoInt & PNN & DCN & FiBiNET & DeepFM & xDeepFM & Avg\\
    \midrule
    
\multirow{4}{*}{\bf \shortstack{Movielens- \\ gender}} & Base & 80.25 & \bf 80.57 & 79.91 & 79.95 & \bf 80.51 & 80.71 & 80.22 & 80.30 & 80.30 \\
     & MLoRA & 80.20 & 80.57 & \bf 80.27 & \bf 80.51 & 80.37 & 80.68 & \bf 80.38 & \bf 80.38 & \bf 80.42 \\
     & MoE-MLoRA & \bf 80.25 & 80.44 & 80.08 & 80.21 & 80.15 & \bf 80.72 & 80.22 & 80.21 & 80.28 \\
     & $\Delta$ & +0.05 & -0.13 & -0.19 & -0.30 & -0.23 & +0.04 & -0.16 & -0.17 & -0.14 \\
\midrule
\multirow{4}{*}{\bf \shortstack{Movielens- \\ age}} & Base & 80.46 & 80.70 & 80.17 & \bf 80.67 & 80.74 & 80.75 & 80.42 & 80.56 & 80.56 \\
     & MLoRA & \bf 80.51 & \bf 80.75 & 80.36 & 80.53 & \bf 80.77 & \bf 80.86 & \bf 80.60 & 80.61 & \bf 80.62 \\
     & MoE-MLoRA & 80.39 & 80.63 & \bf 80.54 & 80.27 & 80.48 & 80.69 & 80.58 & \bf 80.64 & 80.53 \\
     & $\Delta$ & -0.13 & -0.12 & +0.17 & -0.26 & -0.29 & -0.17 & -0.02 & +0.03 & -0.10 \\
\midrule
\multirow{4}{*}{\bf \shortstack{Movielens- \\ occupation}} & Base & 80.53 & 80.49 & 80.59 & 80.40 & \bf 80.60 & 80.53 & 80.43 & \bf 80.62 & 80.52 \\
     & MLoRA & 80.46 & 80.39 & 80.38 & 80.44 & 80.25 & \bf 80.62 & 80.57 & 80.57 & 80.46 \\
     & MoE-MLoRA & \bf 80.54 & \bf 80.55 & \bf 80.59 & \bf 80.60 & 80.57 & 80.59 & \bf 80.59 & 80.56 & \bf 80.58 \\
     & $\Delta$ & +0.08 & +0.16 & +0.21 & +0.16 & +0.32 & -0.04 & +0.02 & -0.01 & +0.11 \\
\midrule
\multirow{4}{*}{\bf \shortstack{Taobao- \\ theme-10}} & Base & 72.42 & 76.21 & 74.19 & \bf 76.92 & 72.02 & 75.83 & 75.19 & 74.71 & 74.69 \\
     & MLoRA & 72.15 & 76.45 & 72.09 & 75.88 & 72.18 & 76.82 & 74.70 & 75.35 & 74.45 \\
     & MoE-MLoRA & \bf 73.22 & \bf 77.19 & \bf 74.89 & 76.67 & \bf 74.47 & \bf 76.91 & \bf 77.16 & \bf 76.57 & \bf 75.89 \\
     & $\Delta$ & +1.07 & +0.74 & +2.80 & +0.79 & +2.28 & +0.08 & +2.46 & +1.22 & +1.43 \\
\midrule
\multirow{4}{*}{\bf \shortstack{Taobao- \\ theme-20}} & Base & 74.74 & 77.40 & 78.16 & 78.79 & \bf 76.54 & 78.99 & 77.12 & 77.16 & 77.36 \\
     & MLoRA & 75.16 & 78.06 & 79.09 & 78.17 & 74.02 & 79.22 & 76.38 & 76.87 & 77.12 \\
     & MoE-MLoRA & \bf 77.00 & \bf 78.22 & \bf 79.66 & \bf 80.71 & 76.35 & \bf 79.75 & \bf 78.93 & \bf 77.92 & \bf 78.57 \\
     & $\Delta$ & +1.84 & +0.16 & +0.57 & +2.54 & +2.33 & +0.54 & +2.55 & +1.04 & +1.45 \\
\midrule
{\bf - } & Avg($\Delta$) & +0.73 & +0.21 & +0.88 & +0.77 & +1.12 & +0.13 & +1.22 & +0.54 & +0.70 \\
\bottomrule
    \end{tabular}
    }
    \end{table*}

\subsection{Effect of the Number of Experts per Domain}

\begin{figure}[H]
    \centering
    \includegraphics[width=0.9\columnwidth]{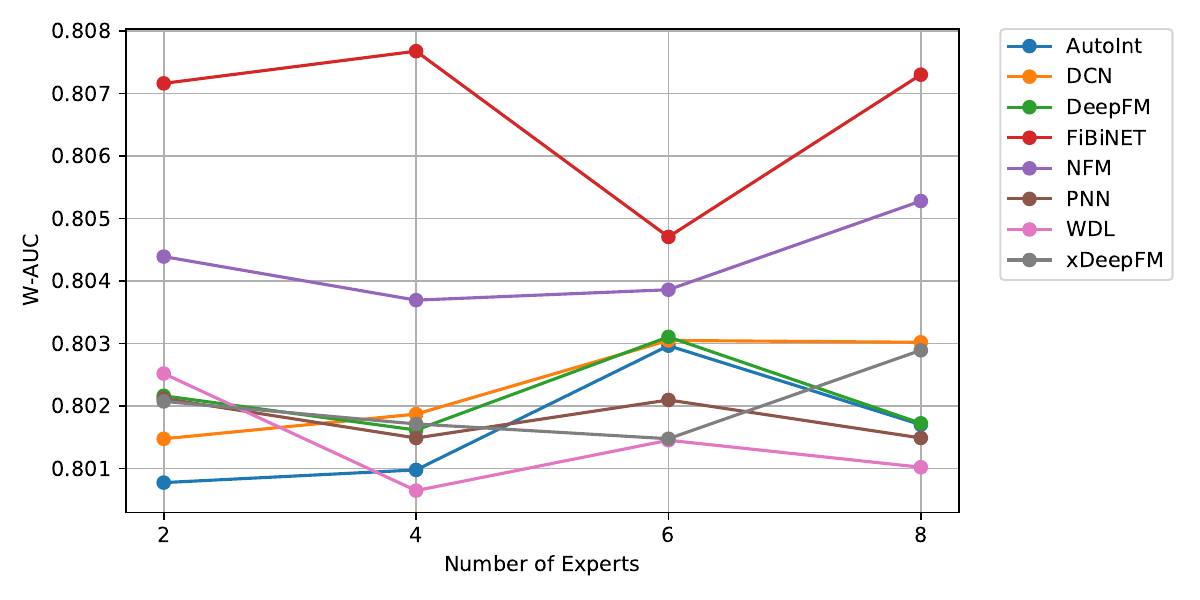}
    \caption{Effect of the number of experts on W-AUC performance across models. Results are based on the Movielens dataset split by gender into two domains.}

    \label{fig:wauc_vs_experts}
\end{figure}

Figure~\ref{fig:wauc_vs_experts} illustrates the WAUC scores as a function of the number of experts across different models.
The results indicate that the impact of the number of experts varies across models, with most architectures showing no clear improvement as the number of experts increases, and performance fluctuating. 
While DeepFM and AutoInt achieve peak performance with a moderate number of experts (6 experts), larger ensembles appear to negatively impact their performance. 
In contrast, models like NFM benefit from a larger ensemble, suggesting that the effect of expert specialization is highly model-dependent.

\section{Discussion}
\label{sec:discussion}

Our results demonstrate that MoE-MLoRA yields strong improvements in complex, sparse, and large-scale datasets like Taobao, but shows limited or inconsistent gains on Movielens. 
The key difference lies in dataset characteristics—while Taobao is sparse and diverse, allowing expert models to specialize effectively, Movielens has dense interactions and limited domain variety, making expert-based adaptation less useful and, in some cases, even counterproductive. 
One possible explanation is that the Movielens splits (gender and age) may not have created sufficiently distinct distributions, meaning that the data across groups remained too similar.
If user behavior and interaction patterns were nearly identical across domains, the gating function in MoE-MLoRA would struggle to assign meaningful expert specializations, leading to minimal gains or even slight performance degradation.

Our second experiment on the number of experts per domain reveals that increasing the number of experts does not consistently improve performance. While some models exhibited slight variations in performance with more experts, the changes were not particularly noticeable. Our hypothesis is that our method is most effective when applied to domains that are distinct but still share transferable knowledge. In this scenario, each expert specializes in a specific aspect, and the gating function effectively combines their contributions, leveraging the strengths of all experts. However, simply increasing the number of experts does not necessarily lead to learning diverse representations. Instead, many of the additional experts become redundant, failing to contribute meaningful new insights and potentially adding unnecessary complexity.



\section{Conclusion}
\label{sec:conclusion}

In this work, we addressed the challenge of multi-domain CTR prediction by proposing MoE-MLoRA, an expert-based adaptation of the MLoRA method.
While MLoRA applies a single LoRA adapter per task, our approach introduces a MoE, allowing different experts to specialize in different domains and finally use a Gating Network to weight their contribution based on the input’s domain context.
The goal was to determine whether expert-based specialization improves predictive performance across multiple recommendation models.

Our experimental evaluation shows that MoE-MLoRA delivers substantial performance gains in large-scale and highly sparse datasets, such as Taobao, where clear domain separation and limited interaction density make expert specialization more beneficial. 
In particular, we observe average WAUC improvements of +1.43 and +1.45 over MLoRA on the Taobao-10 and Taobao-20 splits, respectively. 
These findings underscore that expert-based architectures are particularly effective in settings with well-defined, behaviorally diverse domains—provided there remains some degree of shared structure that enables cross-domain knowledge transfer. 
This has important implications for designing adaptive recommendation models, especially in real-world environments characterized by fragmented user behavior and heterogeneous content domains.




For future work, we propose investigating improved gating mechanisms that can better capture cross-domain relationships and dynamically adjust expert contributions. 
Additionally, incorporating a preprocessing step to enhance expert diversity when multiple experts are used for a single domain could further improve specialization and prevent redundancy. 
These directions could further enhance the adaptability and efficiency of MoE-based models in large-scale, multi-domain recommendation systems.


\bibliographystyle{ACM-Reference-Format}
\bibliography{references}


\begin{thebibliography}{22}


\ifx \showCODEN    \undefined \def \showCODEN     #1{\unskip}     \fi
\ifx \showDOI      \undefined \def \showDOI       #1{#1}\fi
\ifx \showISBNx    \undefined \def \showISBNx     #1{\unskip}     \fi
\ifx \showISBNxiii \undefined \def \showISBNxiii  #1{\unskip}     \fi
\ifx \showISSN     \undefined \def \showISSN      #1{\unskip}     \fi
\ifx \showLCCN     \undefined \def \showLCCN      #1{\unskip}     \fi
\ifx \shownote     \undefined \def \shownote      #1{#1}          \fi
\ifx \showarticletitle \undefined \def \showarticletitle #1{#1}   \fi
\ifx \showURL      \undefined \def \showURL       {\relax}        \fi
\providecommand\bibfield[2]{#2}
\providecommand\bibinfo[2]{#2}
\providecommand\natexlab[1]{#1}
\providecommand\showeprint[2][]{arXiv:#2}

\bibitem[Bian et~al\mbox{.}(2021)]%
        {bian2021canfeaturecoactionclickthrough}
\bibfield{author}{\bibinfo{person}{Weijie Bian}, \bibinfo{person}{Kailun Wu}, \bibinfo{person}{Lejian Ren}, \bibinfo{person}{Qi Pi}, \bibinfo{person}{Yujing Zhang}, \bibinfo{person}{Can Xiao}, \bibinfo{person}{Xiang-Rong Sheng}, \bibinfo{person}{Yong-Nan Zhu}, \bibinfo{person}{Zhangming Chan}, \bibinfo{person}{Na Mou}, \bibinfo{person}{Xinchen Luo}, \bibinfo{person}{Shiming Xiang}, \bibinfo{person}{Guorui Zhou}, \bibinfo{person}{Xiaoqiang Zhu}, {and} \bibinfo{person}{Hongbo Deng}.} \bibinfo{year}{2021}\natexlab{}.
\newblock \bibinfo{title}{CAN: Feature Co-Action for Click-Through Rate Prediction}.
\newblock
\newblock
\showeprint[arxiv]{2011.05625}~[cs.IR]
\urldef\tempurl%
\url{https://arxiv.org/abs/2011.05625}
\showURL{%
\tempurl}


\bibitem[Blondel et~al\mbox{.}(2016)]%
        {10.5555/3157382.3157473}
\bibfield{author}{\bibinfo{person}{Mathieu Blondel}, \bibinfo{person}{Akinori Fujino}, \bibinfo{person}{Naonori Ueda}, {and} \bibinfo{person}{Masakazu Ishihata}.} \bibinfo{year}{2016}\natexlab{}.
\newblock \showarticletitle{Higher-order factorization machines}. In \bibinfo{booktitle}{\emph{Proceedings of the 30th International Conference on Neural Information Processing Systems}} (Barcelona, Spain) \emph{(\bibinfo{series}{NIPS'16})}. \bibinfo{publisher}{Curran Associates Inc.}, \bibinfo{address}{Red Hook, NY, USA}, \bibinfo{pages}{3359–3367}.
\newblock
\showISBNx{9781510838819}


\bibitem[Cheng et~al\mbox{.}(2016)]%
        {wide_and_deep}
\bibfield{author}{\bibinfo{person}{Heng-Tze Cheng}, \bibinfo{person}{Levent Koc}, \bibinfo{person}{Jeremiah Harmsen}, \bibinfo{person}{Tal Shaked}, \bibinfo{person}{Tushar Chandra}, \bibinfo{person}{Hrishi Aradhye}, \bibinfo{person}{Glen Anderson}, \bibinfo{person}{Greg Corrado}, \bibinfo{person}{Wei Chai}, \bibinfo{person}{Mustafa Ispir}, \bibinfo{person}{Rohan Anil}, \bibinfo{person}{Zakaria Haque}, \bibinfo{person}{Lichan Hong}, \bibinfo{person}{Vihan Jain}, \bibinfo{person}{Xiaobing Liu}, {and} \bibinfo{person}{Hemal Shah}.} \bibinfo{year}{2016}\natexlab{}.
\newblock \showarticletitle{Wide \& Deep Learning for Recommender Systems}. In \bibinfo{booktitle}{\emph{Proceedings of the 1st Workshop on Deep Learning for Recommender Systems}} (Boston, MA, USA) \emph{(\bibinfo{series}{DLRS 2016})}. \bibinfo{publisher}{Association for Computing Machinery}, \bibinfo{address}{New York, NY, USA}, \bibinfo{pages}{7–10}.
\newblock
\showISBNx{9781450347952}
\urldef\tempurl%
\url{https://doi.org/10.1145/2988450.2988454}
\showDOI{\tempurl}


\bibitem[Dredze et~al\mbox{.}(2010)]%
        {dredze2010multi}
\bibfield{author}{\bibinfo{person}{Mark Dredze}, \bibinfo{person}{Alex Kulesza}, {and} \bibinfo{person}{Koby Crammer}.} \bibinfo{year}{2010}\natexlab{}.
\newblock \showarticletitle{Multi-domain learning by confidence-weighted parameter combination}.
\newblock \bibinfo{journal}{\emph{Machine Learning}}  \bibinfo{volume}{79} (\bibinfo{year}{2010}), \bibinfo{pages}{123--149}.
\newblock


\bibitem[Du et~al\mbox{.}(2019)]%
        {du2019sequential}
\bibfield{author}{\bibinfo{person}{Zhengxiao Du}, \bibinfo{person}{Xiaowei Wang}, \bibinfo{person}{Hongxia Yang}, \bibinfo{person}{Jingren Zhou}, {and} \bibinfo{person}{Jie Tang}.} \bibinfo{year}{2019}\natexlab{}.
\newblock \showarticletitle{Sequential scenario-specific meta learner for online recommendation}. In \bibinfo{booktitle}{\emph{Proceedings of the 25th ACM SIGKDD International Conference on Knowledge Discovery \& Data Mining}}. \bibinfo{pages}{2895--2904}.
\newblock


\bibitem[Guo et~al\mbox{.}(2017)]%
        {guo2017deepfmfactorizationmachinebasedneural}
\bibfield{author}{\bibinfo{person}{Huifeng Guo}, \bibinfo{person}{Ruiming Tang}, \bibinfo{person}{Yunming Ye}, \bibinfo{person}{Zhenguo Li}, {and} \bibinfo{person}{Xiuqiang He}.} \bibinfo{year}{2017}\natexlab{}.
\newblock \bibinfo{title}{DeepFM: A Factorization-Machine based Neural Network for CTR Prediction}.
\newblock
\newblock
\showeprint[arxiv]{1703.04247}~[cs.IR]
\urldef\tempurl%
\url{https://arxiv.org/abs/1703.04247}
\showURL{%
\tempurl}


\bibitem[He and McAuley(2016)]%
        {he2016ups}
\bibfield{author}{\bibinfo{person}{Ruining He} {and} \bibinfo{person}{Julian McAuley}.} \bibinfo{year}{2016}\natexlab{}.
\newblock \showarticletitle{Ups and downs: Modeling the visual evolution of fashion trends with one-class collaborative filtering}. In \bibinfo{booktitle}{\emph{proceedings of the 25th international conference on world wide web}}. \bibinfo{pages}{507--517}.
\newblock


\bibitem[He and Chua(2017)]%
        {he2017neural}
\bibfield{author}{\bibinfo{person}{Xiangnan He} {and} \bibinfo{person}{Tat-Seng Chua}.} \bibinfo{year}{2017}\natexlab{}.
\newblock \showarticletitle{Neural Factorization Machines for Sparse Predictive Analytics}. In \bibinfo{booktitle}{\emph{Proceedings of the 40th International ACM SIGIR Conference on Research and Development in Information Retrieval}}. \bibinfo{pages}{355--364}.
\newblock


\bibitem[Hu et~al\mbox{.}(2022)]%
        {hu2022lora}
\bibfield{author}{\bibinfo{person}{Edward~J Hu}, \bibinfo{person}{Yelong Shen}, \bibinfo{person}{Phillip Wallis}, \bibinfo{person}{Zeyuan Allen-Zhu}, \bibinfo{person}{Yuanzhi Li}, \bibinfo{person}{Shean Wang}, \bibinfo{person}{Lu Wang}, \bibinfo{person}{Weizhu Chen}, {et~al\mbox{.}}} \bibinfo{year}{2022}\natexlab{}.
\newblock \showarticletitle{Lora: Low-rank adaptation of large language models.}
\newblock \bibinfo{journal}{\emph{ICLR}} \bibinfo{volume}{1}, \bibinfo{number}{2} (\bibinfo{year}{2022}), \bibinfo{pages}{3}.
\newblock


\bibitem[Huang et~al\mbox{.}(2019)]%
        {huang2019fibinet}
\bibfield{author}{\bibinfo{person}{Tao Huang}, \bibinfo{person}{Zhiqiang Zhang}, \bibinfo{person}{Junlin Zhang}, \bibinfo{person}{Yongjun Zhu}, \bibinfo{person}{Rui Xu}, \bibinfo{person}{Jiang Bian}, \bibinfo{person}{Rui Xie}, {and} \bibinfo{person}{Ruoming Jin}.} \bibinfo{year}{2019}\natexlab{}.
\newblock \showarticletitle{FiBiNET: Combining Feature Importance and Bilinear Feature Interaction for Click-Through Rate Prediction}. In \bibinfo{booktitle}{\emph{Proceedings of the 13th ACM Conference on Recommender Systems}}. \bibinfo{pages}{169--177}.
\newblock


\bibitem[Juan et~al\mbox{.}(2016)]%
        {10.1145/2959100.2959134}
\bibfield{author}{\bibinfo{person}{Yuchin Juan}, \bibinfo{person}{Yong Zhuang}, \bibinfo{person}{Wei-Sheng Chin}, {and} \bibinfo{person}{Chih-Jen Lin}.} \bibinfo{year}{2016}\natexlab{}.
\newblock \showarticletitle{Field-aware Factorization Machines for CTR Prediction}. In \bibinfo{booktitle}{\emph{Proceedings of the 10th ACM Conference on Recommender Systems}} (Boston, Massachusetts, USA) \emph{(\bibinfo{series}{RecSys '16})}. \bibinfo{publisher}{Association for Computing Machinery}, \bibinfo{address}{New York, NY, USA}, \bibinfo{pages}{43–50}.
\newblock
\showISBNx{9781450340359}
\urldef\tempurl%
\url{https://doi.org/10.1145/2959100.2959134}
\showDOI{\tempurl}


\bibitem[Li et~al\mbox{.}(2017)]%
        {Li_2017_ICCV}
\bibfield{author}{\bibinfo{person}{Da Li}, \bibinfo{person}{Yongxin Yang}, \bibinfo{person}{Yi-Zhe Song}, {and} \bibinfo{person}{Timothy~M. Hospedales}.} \bibinfo{year}{2017}\natexlab{}.
\newblock \showarticletitle{Deeper, Broader and Artier Domain Generalization}. In \bibinfo{booktitle}{\emph{Proceedings of the IEEE International Conference on Computer Vision (ICCV)}}.
\newblock


\bibitem[Li et~al\mbox{.}(2020)]%
        {li2020improving}
\bibfield{author}{\bibinfo{person}{Pengcheng Li}, \bibinfo{person}{Runze Li}, \bibinfo{person}{Qing Da}, \bibinfo{person}{An-Xiang Zeng}, {and} \bibinfo{person}{Lijun Zhang}.} \bibinfo{year}{2020}\natexlab{}.
\newblock \showarticletitle{Improving multi-scenario learning to rank in e-commerce by exploiting task relationships in the label space}. In \bibinfo{booktitle}{\emph{Proceedings of the 29th ACM International Conference on Information \& Knowledge Management}}. \bibinfo{pages}{2605--2612}.
\newblock


\bibitem[Lian et~al\mbox{.}(2018)]%
        {lian2018xdeepfm}
\bibfield{author}{\bibinfo{person}{Jianxun Lian}, \bibinfo{person}{Xiaohuan Zhou}, \bibinfo{person}{Fuzheng Zhang}, \bibinfo{person}{Zhongxia Chen}, \bibinfo{person}{Xing Xie}, {and} \bibinfo{person}{Guangzhong Sun}.} \bibinfo{year}{2018}\natexlab{}.
\newblock \showarticletitle{xDeepFM: Combining Explicit and Implicit Feature Interactions for Recommender Systems}.
\newblock \bibinfo{journal}{\emph{arXiv preprint arXiv:1803.05170}} (\bibinfo{year}{2018}).
\newblock


\bibitem[Liu et~al\mbox{.}(2023)]%
        {liu2023moelora}
\bibfield{author}{\bibinfo{person}{Qidong Liu}, \bibinfo{person}{Xian Wu}, \bibinfo{person}{Xiangyu Zhao}, \bibinfo{person}{Yuanshao Zhu}, \bibinfo{person}{Derong Xu}, \bibinfo{person}{Feng Tian}, {and} \bibinfo{person}{Yefeng Zheng}.} \bibinfo{year}{2023}\natexlab{}.
\newblock \showarticletitle{Moelora: An moe-based parameter efficient fine-tuning method for multi-task medical applications}.
\newblock \bibinfo{journal}{\emph{CoRR}} (\bibinfo{year}{2023}).
\newblock


\bibitem[Luo et~al\mbox{.}(2024)]%
        {luo2024moelora}
\bibfield{author}{\bibinfo{person}{Tongxu Luo}, \bibinfo{person}{Jiahe Lei}, \bibinfo{person}{Fangyu Lei}, \bibinfo{person}{Weihao Liu}, \bibinfo{person}{Shizhu He}, \bibinfo{person}{Jun Zhao}, {and} \bibinfo{person}{Kang Liu}.} \bibinfo{year}{2024}\natexlab{}.
\newblock \showarticletitle{Moelora: Contrastive learning guided mixture of experts on parameter-efficient fine-tuning for large language models}.
\newblock \bibinfo{journal}{\emph{arXiv preprint arXiv:2402.12851}} (\bibinfo{year}{2024}).
\newblock


\bibitem[Qu et~al\mbox{.}(2016)]%
        {7837964}
\bibfield{author}{\bibinfo{person}{Yanru Qu}, \bibinfo{person}{Han Cai}, \bibinfo{person}{Kan Ren}, \bibinfo{person}{Weinan Zhang}, \bibinfo{person}{Yong Yu}, \bibinfo{person}{Ying Wen}, {and} \bibinfo{person}{Jun Wang}.} \bibinfo{year}{2016}\natexlab{}.
\newblock \showarticletitle{Product-Based Neural Networks for User Response Prediction}. In \bibinfo{booktitle}{\emph{2016 IEEE 16th International Conference on Data Mining (ICDM)}}. \bibinfo{pages}{1149--1154}.
\newblock
\urldef\tempurl%
\url{https://doi.org/10.1109/ICDM.2016.0151}
\showDOI{\tempurl}


\bibitem[Song et~al\mbox{.}(2019)]%
        {AutoInt}
\bibfield{author}{\bibinfo{person}{Weiping Song}, \bibinfo{person}{Chence Shi}, \bibinfo{person}{Zhiping Xiao}, \bibinfo{person}{Zhijian Duan}, \bibinfo{person}{Yewen Xu}, \bibinfo{person}{Ming Zhang}, {and} \bibinfo{person}{Jian Tang}.} \bibinfo{year}{2019}\natexlab{}.
\newblock \showarticletitle{AutoInt: Automatic Feature Interaction Learning via Self-Attentive Neural Networks}. In \bibinfo{booktitle}{\emph{Proceedings of the 28th ACM International Conference on Information and Knowledge Management}} (Beijing, China) \emph{(\bibinfo{series}{CIKM '19})}. \bibinfo{publisher}{Association for Computing Machinery}, \bibinfo{address}{New York, NY, USA}, \bibinfo{pages}{1161–1170}.
\newblock
\showISBNx{9781450369763}
\urldef\tempurl%
\url{https://doi.org/10.1145/3357384.3357925}
\showDOI{\tempurl}


\bibitem[Tang et~al\mbox{.}(2020)]%
        {10.1145/3383313.3412236}
\bibfield{author}{\bibinfo{person}{Hongyan Tang}, \bibinfo{person}{Junning Liu}, \bibinfo{person}{Ming Zhao}, {and} \bibinfo{person}{Xudong Gong}.} \bibinfo{year}{2020}\natexlab{}.
\newblock \showarticletitle{Progressive Layered Extraction (PLE): A Novel Multi-Task Learning (MTL) Model for Personalized Recommendations}. In \bibinfo{booktitle}{\emph{Proceedings of the 14th ACM Conference on Recommender Systems}} (Virtual Event, Brazil) \emph{(\bibinfo{series}{RecSys '20})}. \bibinfo{publisher}{Association for Computing Machinery}, \bibinfo{address}{New York, NY, USA}, \bibinfo{pages}{269–278}.
\newblock
\showISBNx{9781450375832}
\urldef\tempurl%
\url{https://doi.org/10.1145/3383313.3412236}
\showDOI{\tempurl}


\bibitem[Wang et~al\mbox{.}(2017)]%
        {wang2017deep}
\bibfield{author}{\bibinfo{person}{Ruoxi Wang}, \bibinfo{person}{Bin Fu}, \bibinfo{person}{Gang Fu}, {and} \bibinfo{person}{Mingliang Wang}.} \bibinfo{year}{2017}\natexlab{}.
\newblock \showarticletitle{Deep \& Cross Network for Ad Click Predictions}. In \bibinfo{booktitle}{\emph{Proceedings of the ADKDD'17}}. \bibinfo{pages}{1--7}.
\newblock


\bibitem[Yang et~al\mbox{.}(2024)]%
        {mlora_recsys2024}
\bibfield{author}{\bibinfo{person}{Zhiming Yang}, \bibinfo{person}{Haining Gao}, \bibinfo{person}{Dehong Gao}, \bibinfo{person}{Luwei Yang}, \bibinfo{person}{Libin Yang}, \bibinfo{person}{Xiaoyan Cai}, \bibinfo{person}{Wei Ning}, {and} \bibinfo{person}{Guannan Zhang}.} \bibinfo{year}{2024}\natexlab{}.
\newblock \showarticletitle{MLoRA: Multi-Domain Low-Rank Adaptive Network for CTR Prediction}. In \bibinfo{booktitle}{\emph{Proceedings of the 18th ACM Conference on Recommender Systems}} (Bari, Italy) \emph{(\bibinfo{series}{RecSys '24})}. \bibinfo{publisher}{Association for Computing Machinery}, \bibinfo{address}{New York, NY, USA}, \bibinfo{pages}{287–297}.
\newblock
\showISBNx{9798400705052}
\urldef\tempurl%
\url{https://doi.org/10.1145/3640457.3688134}
\showDOI{\tempurl}


\bibitem[Zhou et~al\mbox{.}(2023)]%
        {9847099}
\bibfield{author}{\bibinfo{person}{Kaiyang Zhou}, \bibinfo{person}{Ziwei Liu}, \bibinfo{person}{Yu Qiao}, \bibinfo{person}{Tao Xiang}, {and} \bibinfo{person}{Chen~Change Loy}.} \bibinfo{year}{2023}\natexlab{}.
\newblock \showarticletitle{Domain Generalization: A Survey}.
\newblock \bibinfo{journal}{\emph{IEEE Transactions on Pattern Analysis and Machine Intelligence}} \bibinfo{volume}{45}, \bibinfo{number}{4} (\bibinfo{year}{2023}), \bibinfo{pages}{4396--4415}.
\newblock
\urldef\tempurl%
\url{https://doi.org/10.1109/TPAMI.2022.3195549}
\showDOI{\tempurl}


\end{thebibliography}

\end{document}